\documentstyle[editedvolume]{crckapb} 
\def\PsfigVersion{1.10}
\def\setDriver{\DvipsDriver} 
\ifx\undefined\psfig\else\endinput\fi
\let\LaTeXAtSign=\@
\let\@=\relax
\edef\psfigRestoreAt{\catcode`\@=\number\catcode`@\relax}
\catcode`\@=11\relax
\newwrite\@unused
\def\ps@typeout#1{{\let\protect\string\immediate\write\@unused{#1}}}

\def\DvipsDriver{
	\ps@typeout{psfig/tex \PsfigVersion -dvips}
\def\PsfigSpecials{\DvipsSpecials} 	\def\ps@dir{/}
\def\ps@predir{} }
\def\OzTeXDriver{
	\ps@typeout{psfig/tex \PsfigVersion -oztex}
	\def\PsfigSpecials{\OzTeXSpecials}
	\def\ps@dir{:}
	\def\ps@predir{:}
	\catcode`\^^J=5
}

\def\figurepath{./:}

\def\DoPaths#1{\expandafter\EachPath#1\stoplist}
\def\leer{}
\def\EachPath#1:#2\stoplist{
  \ExistsFile{#1}{\SearchedFile}
  \ifx#2\leer
  \else
    \expandafter\EachPath#2\stoplist
  \fi}
\def\ps@dir{/}
\def\ExistsFile#1#2{%
   \openin1=\ps@predir#1\ps@dir#2
   \ifeof1
       \closein1
   \else
       \closein1
        \ifx\ps@founddir\leer
           \edef\ps@founddir{#1}
        \fi
   \fi}
\def\get@dir#1{%
  \def\ps@founddir{}
  \def\SearchedFile{#1}
  \DoPaths\figurepath
}

\def\@nnil{\@nil}
\def\@empty{}
\def\@psdonoop#1\@@#2#3{}
\def\@psdo#1:=#2\do#3{\edef\@psdotmp{#2}\ifx\@psdotmp\@empty \else
    \expandafter\@psdoloop#2,\@nil,\@nil\@@#1{#3}\fi}
\def\@psdoloop#1,#2,#3\@@#4#5{\def#4{#1}\ifx #4\@nnil \else
       #5\def#4{#2}\ifx #4\@nnil \else#5\@ipsdoloop #3\@@#4{#5}\fi\fi}
\def\@ipsdoloop#1,#2\@@#3#4{\def#3{#1}\ifx #3\@nnil 
       \let\@nextwhile=\@psdonoop \else
      #4\relax\let\@nextwhile=\@ipsdoloop\fi\@nextwhile#2\@@#3{#4}}
\def\@tpsdo#1:=#2\do#3{\xdef\@psdotmp{#2}\ifx\@psdotmp\@empty \else
    \@tpsdoloop#2\@nil\@nil\@@#1{#3}\fi}
\def\@tpsdoloop#1#2\@@#3#4{\def#3{#1}\ifx #3\@nnil 
       \let\@nextwhile=\@psdonoop \else
      #4\relax\let\@nextwhile=\@tpsdoloop\fi\@nextwhile#2\@@#3{#4}}
\ifx\undefined\fbox
\newdimen\fboxrule
\newdimen\fboxsep
\newdimen\ps@tempdima
\newbox\ps@tempboxa
\long\def\fbox#1{\leavevmode\setbox\ps@tempboxa\hbox{#1}\ps@tempdima\fboxrule
    \advance\ps@tempdima \fboxsep \advance\ps@tempdima \dp\ps@tempboxa
   \hbox{\lower \ps@tempdima\hbox
  {\vbox{\hrule height \fboxrule
          \hbox{\vrule width \fboxrule \hskip\fboxsep
          \vbox{\vskip\fboxsep \box\ps@tempboxa\vskip\fboxsep}\hskip 
                 \fboxsep\vrule width \fboxrule}
                 \hrule height \fboxrule}}}}
\fi
\newread\ps@stream
\newif\ifnot@eof       
\newif\if@noisy        
\newif\if@atend        
\newif\if@psfile       
{\catcode`\%=12\global\gdef\epsf@start{
\def\epsf@PS{PS}
\def\epsf@getbb#1{%
\openin\ps@stream=\ps@predir#1
\ifeof\ps@stream\ps@typeout{Error, File #1 not found}\else
   {\not@eoftrue \chardef\other=12
    \def\do##1{\catcode`##1=\other}\dospecials \catcode`\ =10
    \loop
       \if@psfile
	  \read\ps@stream to \epsf@fileline
       \else{
	  \obeyspaces
          \read\ps@stream to \epsf@tmp\global\let\epsf@fileline\epsf@tmp}
       \fi
       \ifeof\ps@stream\not@eoffalse\else
       \if@psfile\else
       \expandafter\epsf@test\epsf@fileline:. \\%
       \fi
          \expandafter\epsf@aux\epsf@fileline:. \\%
       \fi
   \ifnot@eof\repeat
   }\closein\ps@stream\fi}%
\long\def\epsf@test#1#2#3:#4\\{\def\epsf@testit{#1#2}
			\ifx\epsf@testit\epsf@start\else
\ps@typeout{Warning! File does not start with `\epsf@start'.  It may not be a PostScript file.}
			\fi
			\@psfiletrue} 
{\catcode`\%=12\global\let\epsf@percent=
\long\def\epsf@aux#1#2:#3\\{\ifx#1\epsf@percent
   \def\epsf@testit{#2}\ifx\epsf@testit\epsf@bblit
	\@atendfalse
        \epsf@atend #3 . \\%
	\if@atend	
	   \if@verbose{
		\ps@typeout{psfig: found `(atend)'; continuing search}
	   }\fi
        \else
        \epsf@grab #3 . . . \\%
        \not@eoffalse
        \global\no@bbfalse
        \fi
   \fi\fi}%
\def\epsf@grab #1 #2 #3 #4 #5\\{%
   \global\def\epsf@llx{#1}\ifx\epsf@llx\empty
      \epsf@grab #2 #3 #4 #5 .\\\else
   \global\def\epsf@lly{#2}%
   \global\def\epsf@urx{#3}\global\def\epsf@ury{#4}\fi}%
\def\epsf@atendlit{(atend)} 
\def\epsf@atend #1 #2 #3\\{%
   \def\epsf@tmp{#1}\ifx\epsf@tmp\empty
      \epsf@atend #2 #3 .\\\else
   \ifx\epsf@tmp\epsf@atendlit\@atendtrue\fi\fi}

\chardef\psletter = 11 
\chardef\other = 12

\newif \ifdebug 
\newif\ifc@mpute 
\c@mputetrue 

\let\then = \relax
\def\r@dian{pt }
\let\r@dians = \r@dian
\let\dimensionless@nit = \r@dian
\let\dimensionless@nits = \dimensionless@nit
\def\internal@nit{sp }
\let\internal@nits = \internal@nit
\newif\ifstillc@nverging
\def \Mess@ge #1{\ifdebug \then \message {#1} \fi}

{ 
	\catcode `\@ = \psletter
	\gdef \nodimen {\expandafter \n@dimen \the \dimen}
	\gdef \term #1 #2 #3%
	       {\edef \t@ {\the #1}
		\edef \t@@ {\expandafter \n@dimen \the #2\r@dian}%
		\t@rm {\t@} {\t@@} {#3}%
	       }
	\gdef \t@rm #1 #2 #3%
	       {{%
		\count 0 = 0
		\dimen 0 = 1 \dimensionless@nit
		\dimen 2 = #2\relax
		\Mess@ge {Calculating term #1 of \nodimen 2}%
		\loop
		\ifnum	\count 0 < #1
		\then	\advance \count 0 by 1
			\Mess@ge {Iteration \the \count 0 \space}%
			\Multiply \dimen 0 by {\dimen 2}%
			\Mess@ge {After multiplication, term = \nodimen 0}%
			\Divide \dimen 0 by {\count 0}%
			\Mess@ge {After division, term = \nodimen 0}%
		\repeat
		\Mess@ge {Final value for term #1 of 
				\nodimen 2 \space is \nodimen 0}%
		\xdef \Term {#3 = \nodimen 0 \r@dians}%
		\aftergroup \Term
	       }}
	\catcode `\p = \other
	\catcode `\t = \other
	\gdef \n@dimen #1pt{#1} 
}

\def \Divide #1by #2{\divide #1 by #2} 

\def \Multiply #1by #2
       {{
	\count 0 = #1\relax
	\count 2 = #2\relax
	\count 4 = 65536
	\Mess@ge {Before scaling, count 0 = \the \count 0 \space and
			count 2 = \the \count 2}%
	\ifnum	\count 0 > 32767 
	\then	\divide \count 0 by 4
		\divide \count 4 by 4
	\else	\ifnum	\count 0 < -32767
		\then	\divide \count 0 by 4
			\divide \count 4 by 4
		\else
		\fi
	\fi
	\ifnum	\count 2 > 32767 
	\then	\divide \count 2 by 4
		\divide \count 4 by 4
	\else	\ifnum	\count 2 < -32767
		\then	\divide \count 2 by 4
			\divide \count 4 by 4
		\else
		\fi
	\fi
	\multiply \count 0 by \count 2
	\divide \count 0 by \count 4
	\xdef \product {#1 = \the \count 0 \internal@nits}%
	\aftergroup \product
       }}

\def\r@duce{\ifdim\dimen0 > 90\r@dian \then   
		\multiply\dimen0 by -1
		\advance\dimen0 by 180\r@dian
		\r@duce
	    \else \ifdim\dimen0 < -90\r@dian \then  
		\advance\dimen0 by 360\r@dian
		\r@duce
		\fi
	    \fi}

\def\Sine#1%
       {{%
	\dimen 0 = #1 \r@dian
	\r@duce
	\ifdim\dimen0 = -90\r@dian \then
	   \dimen4 = -1\r@dian
	   \c@mputefalse
	\fi
	\ifdim\dimen0 = 90\r@dian \then
	   \dimen4 = 1\r@dian
	   \c@mputefalse
	\fi
	\ifdim\dimen0 = 0\r@dian \then
	   \dimen4 = 0\r@dian
	   \c@mputefalse
	\fi
	\ifc@mpute \then
		\divide\dimen0 by 180
		\dimen0=3.141592654\dimen0
		\dimen 2 = 3.1415926535897963\r@dian 
		\divide\dimen 2 by 2 
		\Mess@ge {Sin: calculating Sin of \nodimen 0}%
		\count 0 = 1 
		\dimen 2 = 1 \r@dian 
		\dimen 4 = 0 \r@dian 
		\loop
			\ifnum	\dimen 2 = 0 
			\then	\stillc@nvergingfalse 
			\else	\stillc@nvergingtrue
			\fi
			\ifstillc@nverging 
			\then	\term {\count 0} {\dimen 0} {\dimen 2}%
				\advance \count 0 by 2
				\count 2 = \count 0
				\divide \count 2 by 2
				\ifodd	\count 2 
				\then	\advance \dimen 4 by \dimen 2
				\else	\advance \dimen 4 by -\dimen 2
				\fi
		\repeat
	\fi		
			\xdef \sine {\nodimen 4}%
       }}

\def\Cosine#1{\ifx\sine\UnDefined\edef\Savesine{\relax}\else
		             \edef\Savesine{\sine}\fi
	{\dimen0=#1\r@dian\advance\dimen0 by 90\r@dian
	 \Sine{\nodimen 0}
	 \xdef\cosine{\sine}
	 \xdef\sine{\Savesine}}}	      

\def\psdraft{
	\def\@psdraft{0}
}
\def\psfull{
	\def\@psdraft{100}
}

\psfull

\newif\if@scalefirst
\def\psscalefirst{\@scalefirsttrue}
\def\psrotatefirst{\@scalefirstfalse}
\psrotatefirst

\newif\if@draftbox
\def\psnodraftbox{
	\@draftboxfalse
}
\def\psdraftbox{
	\@draftboxtrue
}
\@draftboxtrue

\newif\if@prologfile
\newif\if@postlogfile
\def\pssilent{
	\@noisyfalse
}
\def\psnoisy{
	\@noisytrue
}
\psnoisy
\newif\if@bbllx
\newif\if@bblly
\newif\if@bburx
\newif\if@bbury
\newif\if@height
\newif\if@width
\newif\if@rheight
\newif\if@rwidth
\newif\if@angle
\newif\if@clip
\newif\if@verbose
\def\@p@@sclip#1{\@cliptrue}
\newif\if@decmpr
\def\@p@@sfigure#1{\def\@p@sfile{null}\def\@p@sbbfile{null}\@decmprfalse
   \openin1=\ps@predir#1
   \ifeof1
	\closein1
	\get@dir{#1}
	\ifx\ps@founddir\leer
		\openin1=\ps@predir#1.bb
		\ifeof1
			\closein1
			\get@dir{#1.bb}
			\ifx\ps@founddir\leer
				\ps@typeout{Can't find #1 in \figurepath}
			\else
				\@decmprtrue
				\def\@p@sfile{\ps@founddir\ps@dir#1}
				\def\@p@sbbfile{\ps@founddir\ps@dir#1.bb}
			\fi
		\else
			\closein1
			\@decmprtrue
			\def\@p@sfile{#1}
			\def\@p@sbbfile{#1.bb}
		\fi
	\else
		\def\@p@sfile{\ps@founddir\ps@dir#1}
		\def\@p@sbbfile{\ps@founddir\ps@dir#1}
	\fi
   \else
	\closein1
	\def\@p@sfile{#1}
	\def\@p@sbbfile{#1}
   \fi
}
\def\@p@@sfile#1{\@p@@sfigure{#1}}
\def\@p@@sbbllx#1{
		\@bbllxtrue
		\dimen100=#1
		\edef\@p@sbbllx{\number\dimen100}
}
\def\@p@@sbblly#1{
		\@bbllytrue
		\dimen100=#1
		\edef\@p@sbblly{\number\dimen100}
}
\def\@p@@sbburx#1{
		\@bburxtrue
		\dimen100=#1
		\edef\@p@sbburx{\number\dimen100}
}
\def\@p@@sbbury#1{
		\@bburytrue
		\dimen100=#1
		\edef\@p@sbbury{\number\dimen100}
}
\def\@p@@sheight#1{
		\@heighttrue
		\dimen100=#1
   		\edef\@p@sheight{\number\dimen100}
}
\def\@p@@swidth#1{
		\@widthtrue
		\dimen100=#1
		\edef\@p@swidth{\number\dimen100}
}
\def\@p@@srheight#1{
		\@rheighttrue
		\dimen100=#1
		\edef\@p@srheight{\number\dimen100}
}
\def\@p@@srwidth#1{
		\@rwidthtrue
		\dimen100=#1
		\edef\@p@srwidth{\number\dimen100}
}
\def\@p@@sangle#1{
		\@angletrue
		\edef\@p@sangle{#1} 
}
\def\@p@@ssilent#1{ 
		\@verbosefalse
}
\def\@p@@sprolog#1{\@prologfiletrue\def\@prologfileval{#1}}
\def\@p@@spostlog#1{\@postlogfiletrue\def\@postlogfileval{#1}}
\def\@cs@name#1{\csname #1\endcsname}
\def\@setparms#1=#2,{\@cs@name{@p@@s#1}{#2}}
\def\ps@init@parms{
		\@bbllxfalse \@bbllyfalse
		\@bburxfalse \@bburyfalse
		\@heightfalse \@widthfalse
		\@rheightfalse \@rwidthfalse
		\def\@p@sbbllx{}\def\@p@sbblly{}
		\def\@p@sbburx{}\def\@p@sbbury{}
		\def\@p@sheight{}\def\@p@swidth{}
		\def\@p@srheight{}\def\@p@srwidth{}
		\def\@p@sangle{0}
		\def\@p@sfile{} \def\@p@sbbfile{}
		\def\@p@scost{10}
		\def\@sc{}
		\@prologfilefalse
		\@postlogfilefalse
		\@clipfalse
		\if@noisy
			\@verbosetrue
		\else
			\@verbosefalse
		\fi
}
\def\parse@ps@parms#1{
	 	\@psdo\@psfiga:=#1\do
		   {\expandafter\@setparms\@psfiga,}}
\newif\ifno@bb
\def\bb@missing{
	\if@verbose{
		\ps@typeout{psfig: searching \@p@sbbfile \space  for bounding box}
	}\fi
	\no@bbtrue
	\epsf@getbb{\@p@sbbfile}
        \ifno@bb \else \bb@cull\epsf@llx\epsf@lly\epsf@urx\epsf@ury\fi
}	
\def\bb@cull#1#2#3#4{
	\dimen100=#1 bp\edef\@p@sbbllx{\number\dimen100}
	\dimen100=#2 bp\edef\@p@sbblly{\number\dimen100}
	\dimen100=#3 bp\edef\@p@sbburx{\number\dimen100}
	\dimen100=#4 bp\edef\@p@sbbury{\number\dimen100}
	\no@bbfalse
}
\newdimen\p@intvaluex
\newdimen\p@intvaluey
\def\rotate@#1#2{{\dimen0=#1 sp\dimen1=#2 sp
		  \global\p@intvaluex=\cosine\dimen0
		  \dimen3=\sine\dimen1
		  \global\advance\p@intvaluex by -\dimen3
		  \global\p@intvaluey=\sine\dimen0
		  \dimen3=\cosine\dimen1
		  \global\advance\p@intvaluey by \dimen3
		  }}
\def\compute@bb{
		\no@bbfalse
		\if@bbllx \else \no@bbtrue \fi
		\if@bblly \else \no@bbtrue \fi
		\if@bburx \else \no@bbtrue \fi
		\if@bbury \else \no@bbtrue \fi
		\ifno@bb \bb@missing \fi
		\ifno@bb \ps@typeout{FATAL ERROR: no bb supplied or found}
			\no-bb-error
		\fi
		\count203=\@p@sbburx
		\count204=\@p@sbbury
		\advance\count203 by -\@p@sbbllx
		\advance\count204 by -\@p@sbblly
		\edef\ps@bbw{\number\count203}
		\edef\ps@bbh{\number\count204}
		\if@angle 
			\Sine{\@p@sangle}\Cosine{\@p@sangle}
	        	{\dimen100=\maxdimen\xdef\r@p@sbbllx{\number\dimen100}
					    \xdef\r@p@sbblly{\number\dimen100}
			                    \xdef\r@p@sbburx{-\number\dimen100}
					    \xdef\r@p@sbbury{-\number\dimen100}}
                        \def\minmaxtest{
			   \ifnum\number\p@intvaluex<\r@p@sbbllx
			      \xdef\r@p@sbbllx{\number\p@intvaluex}\fi
			   \ifnum\number\p@intvaluex>\r@p@sbburx
			      \xdef\r@p@sbburx{\number\p@intvaluex}\fi
			   \ifnum\number\p@intvaluey<\r@p@sbblly
			      \xdef\r@p@sbblly{\number\p@intvaluey}\fi
			   \ifnum\number\p@intvaluey>\r@p@sbbury
			      \xdef\r@p@sbbury{\number\p@intvaluey}\fi
			   }
			\rotate@{\@p@sbbllx}{\@p@sbblly}
			\minmaxtest
			\rotate@{\@p@sbbllx}{\@p@sbbury}
			\minmaxtest
			\rotate@{\@p@sbburx}{\@p@sbblly}
			\minmaxtest
			\rotate@{\@p@sbburx}{\@p@sbbury}
			\minmaxtest
			\edef\@p@sbbllx{\r@p@sbbllx}\edef\@p@sbblly{\r@p@sbblly}
			\edef\@p@sbburx{\r@p@sbburx}\edef\@p@sbbury{\r@p@sbbury}
		\fi
		\count203=\@p@sbburx
		\count204=\@p@sbbury
		\advance\count203 by -\@p@sbbllx
		\advance\count204 by -\@p@sbblly
		\edef\@bbw{\number\count203}
		\edef\@bbh{\number\count204}
}
\def\in@hundreds#1#2#3{\count240=#2 \count241=#3
		     \count100=\count240	
		     \divide\count100 by \count241
		     \count101=\count100
		     \multiply\count101 by \count241
		     \advance\count240 by -\count101
		     \multiply\count240 by 10
		     \count101=\count240	
		     \divide\count101 by \count241
		     \count102=\count101
		     \multiply\count102 by \count241
		     \advance\count240 by -\count102
		     \multiply\count240 by 10
		     \count102=\count240	
		     \divide\count102 by \count241
		     \count200=#1\count205=0
		     \count201=\count200
			\multiply\count201 by \count100
		 	\advance\count205 by \count201
		     \count201=\count200
			\divide\count201 by 10
			\multiply\count201 by \count101
			\advance\count205 by \count201
		     \count201=\count200
			\divide\count201 by 100
			\multiply\count201 by \count102
			\advance\count205 by \count201
		     \edef\@result{\number\count205}
}
\def\compute@wfromh{
		\in@hundreds{\@p@sheight}{\@bbw}{\@bbh}
		\edef\@p@swidth{\@result}
}
\def\compute@hfromw{
	        \in@hundreds{\@p@swidth}{\@bbh}{\@bbw}
		\edef\@p@sheight{\@result}
}
\def\compute@handw{
		\if@height 
			\if@width
			\else
				\compute@wfromh
			\fi
		\else 
			\if@width
				\compute@hfromw
			\else
				\edef\@p@sheight{\@bbh}
				\edef\@p@swidth{\@bbw}
			\fi
		\fi
}
\def\compute@resv{
		\if@rheight \else \edef\@p@srheight{\@p@sheight} \fi
		\if@rwidth \else \edef\@p@srwidth{\@p@swidth} \fi
}
\def\compute@sizes{
	\compute@bb
	\if@scalefirst\if@angle
	\if@width
	   \in@hundreds{\@p@swidth}{\@bbw}{\ps@bbw}
	   \edef\@p@swidth{\@result}
	\fi
	\if@height
	   \in@hundreds{\@p@sheight}{\@bbh}{\ps@bbh}
	   \edef\@p@sheight{\@result}
	\fi
	\fi\fi
	\compute@handw
	\compute@resv}
\def\OzTeXSpecials{
	\special{empty.ps /@isp {true} def}
	\special{empty.ps \@p@swidth \space \@p@sheight \space
			\@p@sbbllx \space \@p@sbblly \space
			\@p@sbburx \space \@p@sbbury \space
			startTexFig \space }
	\if@clip{
		\if@verbose{
			\ps@typeout{(clip)}
		}\fi
		\special{empty.ps doclip \space }
	}\fi
	\if@angle{
		\if@verbose{
			\ps@typeout{(rotate)}
		}\fi
		\special {empty.ps \@p@sangle \space rotate \space} 
	}\fi
	\if@prologfile
	    \special{\@prologfileval \space } \fi
	\if@decmpr{
		\if@verbose{
			\ps@typeout{psfig: Compression not available
			in OzTeX version \space }
		}\fi
	}\else{
		\if@verbose{
			\ps@typeout{psfig: including \@p@sfile \space }
		}\fi
		\special{epsf=\ps@predir\@p@sfile \space }
	}\fi
	\if@postlogfile
	    \special{\@postlogfileval \space } \fi
	\special{empty.ps /@isp {false} def}
}
\def\DvipsSpecials{
	\special{ps::[begin] 	\@p@swidth \space \@p@sheight \space
			\@p@sbbllx \space \@p@sbblly \space
			\@p@sbburx \space \@p@sbbury \space
			startTexFig \space }
	\if@clip{
		\if@verbose{
			\ps@typeout{(clip)}
		}\fi
		\special{ps:: doclip \space }
	}\fi
	\if@angle
		\if@verbose{
			\ps@typeout{(clip)}
		}\fi
		\special {ps:: \@p@sangle \space rotate \space} 
	\fi
	\if@prologfile
	    \special{ps: plotfile \@prologfileval \space } \fi
	\if@decmpr{
		\if@verbose{
			\ps@typeout{psfig: including \@p@sfile.Z \space }
		}\fi
		\special{ps: plotfile "`zcat \@p@sfile.Z" \space }
	}\else{
		\if@verbose{
			\ps@typeout{psfig: including \@p@sfile \space }
		}\fi
		\special{ps: plotfile \@p@sfile \space }
	}\fi
	\if@postlogfile
	    \special{ps: plotfile \@postlogfileval \space } \fi
	\special{ps::[end] endTexFig \space }
}
\def\psfig#1{\vbox {
	\ps@init@parms
	\parse@ps@parms{#1}
	\compute@sizes
	\ifnum\@p@scost<\@psdraft{
		\PsfigSpecials 
		\vbox to \@p@srheight sp{
			\hbox to \@p@srwidth sp{
				\hss
			}
		\vss
		}
	}\else{
		\if@draftbox{		
			\hbox{\fbox{\vbox to \@p@srheight sp{
			\vss
			\hbox to \@p@srwidth sp{ \hss 
			 \hss }
			\vss
			}}}
		}\else{
			\vbox to \@p@srheight sp{
			\vss
			\hbox to \@p@srwidth sp{\hss}
			\vss
			}
		}\fi

	}\fi
}}
\psfigRestoreAt
\setDriver
\let\@=\LaTeXAtSign

\newcommand{\stt}{\small\tt}

\begin{opening}
\title{ANALYSIS OF THE TYPE IA SUPERNOVA SN1994D} 

\author{P. H\"oflich}
\institute{Harvard University, Center for Astrophysics\\
           60 Garden Str.,  Cambridge, MA 02138, USA}

\end{opening}

\runningtitle{ANALYSIS OF THE TYPE IA SUPERNOVA SN1994D}

\begin{document}
%
%
%
\section{Introduction}
On Mar. 7, 1994,
Treffers, Filippenko and Van Dyk (1994) discovered the Type Ia
 supernova (SN~Ia) 1994D   in the SO   galaxy NGC 4526,
a member of the Virgo cluster which was
 was about 12 days before maximum light.
 Subsequently, both the light curve in Johnson's UBVRI colors
 and spectra a have been monitored by several groups
 (e.g. Filippenko et al. private communication, Smith et al., 1995) 
making this supernova one of the best observed until now.
In addition, starting at Mar 12, IR-spectra are reported to be taken 
 at the Royal Greenwich Observatory
showing a  P-Cygni line at about 1.0 $\mu m$ which, likely, is due to 
He I (Meikle 1995). 

 It is generally accepted that SN~Ia are
  thermonuclear explosions of carbon-oxygen white dwarfs (WD)
(Hoyle \& Fowler 1960). However, details of the scenario are still under debate. 
 For discussions of various theoretical aspects see
e.g. Wheeler \& Harkness 1990, Canal 1995,
Nomoto et al. 1995, H\"oflich \& Khokhlov 1995, Woosley \& Weaver 1995). 
 What we observe as a supernova event is not the explosion itself but the light 
emitted from a rapidly expanding envelope produced by the stellar explosion. As 
the photosphere recedes, deeper layers of the ejecta become visible. A detailed 
analysis of the light curves (LC) and spectra gives us the opportunity to determine the
density, velocity and composition structure of the ejecta.
 
\section{Comparison of light curves and spectra}
 Our analysis is based on  observations of SN1994D   by the 
 supernova group at the Center for Astrophysics.
 For the first time, a detailed  analysis of a Type Ia supernova is presented which 
is consistent both with respect to the explosion mechanism, the optical and infrared light
curves, and the spectral evolution.
 The only free parameters are the initial structure of the WD
and  the description of the burning front.
 Our approach  provides  a direct  link between observational properties 
and both the explosion mechanism and, maybe, the progenitor evolution.
\begin{figure}
\psfig{figure=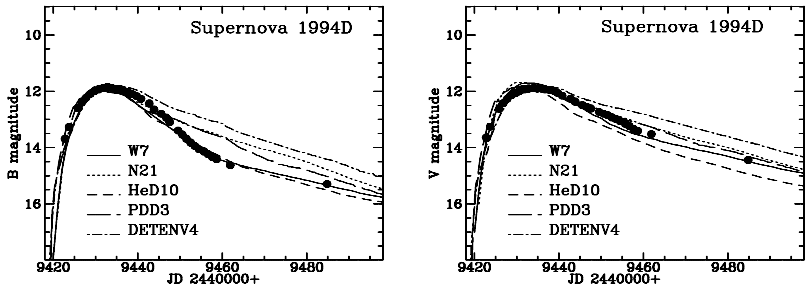,width=12.4cm,angle=270}
\caption{
Observed V and B LCs for SN1994D in comparison to
the theoretical LCs of the deflagration W7, the delayed detonation
N21, the helium detonation HeD10, 
 the pulsating delayed detonation  PDD3, and the envelope model DET2env4.}
\end{figure}
 
 The explosions are calculated using   one-dimensional Lagrangian
hydro with artificial viscosity (Khokhlov, 1991) and radiation-hydro codes (H\"oflich \&
Khokhlov 1995)
including  a nuclear network. Subsequently, 
the LC are constructed. Spectra are computed      
for several instants of time using the density, chemical, and luminosity
structure resulting from the light curve code which includes the solution
of the radiation transport implicitly via the moment equations, expansion opacities,
a detailed Monte-Carlo $\gamma-$ray transport, and a detailed equation of state.
 Our NLTE code for calculating synthetic spectra
 solves the relativistic radiation transport equations in
 comoving frame  consistently  with the
statistical equations and ionization due to $\gamma $ radiation
for the most important elements (He, C, O, Ne, Na, Mg, Si, S, Ca, Fe).
About   300,000 additional lines  are included
 assuming LTE-level populations and an equivalent-two-level approach for the 
source functions. For more details on technical aspects and a discussion of
the relation between observational properties and the underlying model,
 see Khokhlov et al. (1993), H\"oflich (1995), H\"oflich \& Khokhlov 
(1995), H\"oflich et al. (1995, this volume) and references therein.

\begin{figure}
\psfig{figure=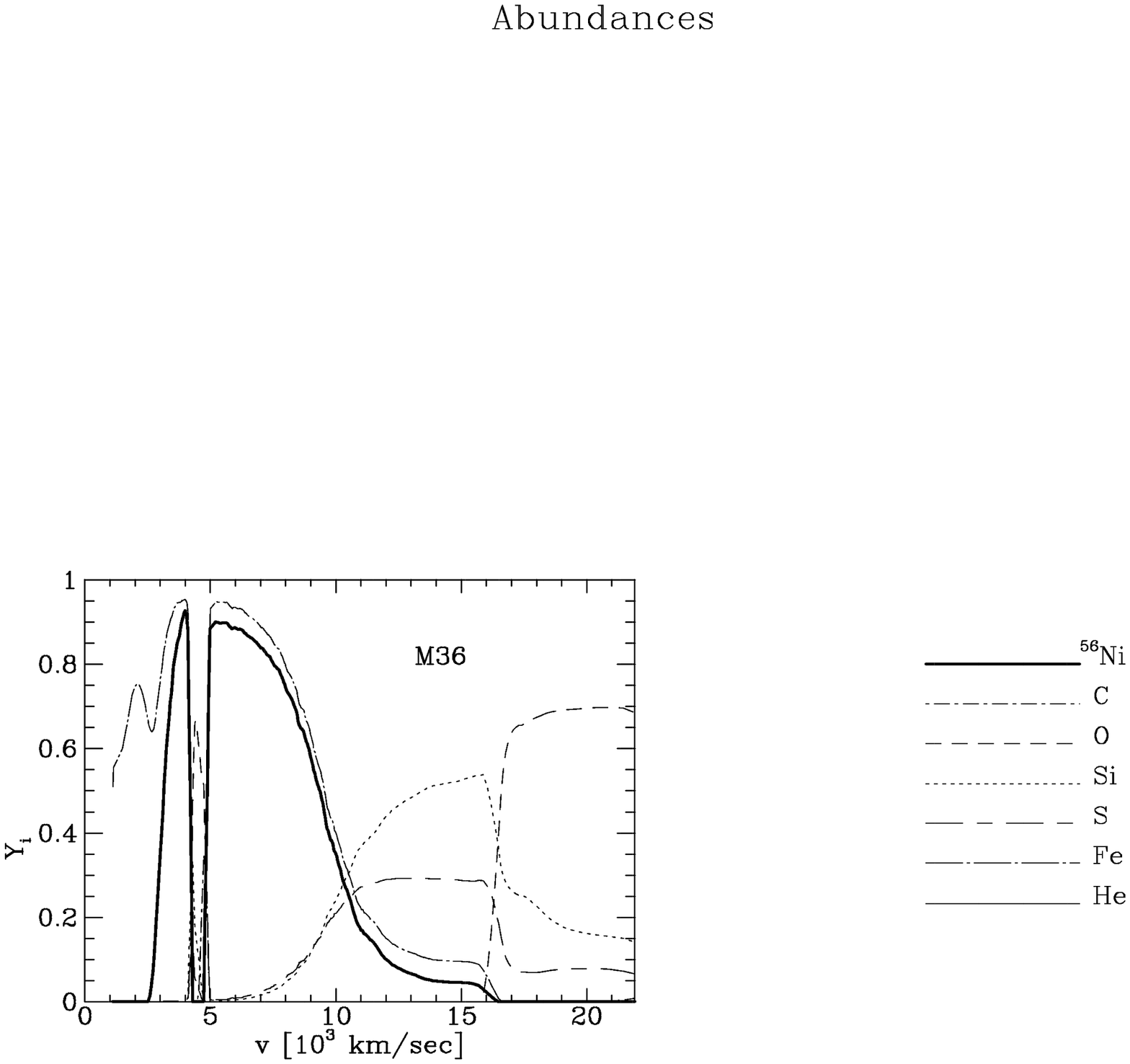,width=12.4cm,angle=270}
\caption{Some Abundances  as a function of the expansion
velocity for the delayed detonation model M36.}
\end{figure}
\begin{figure}
\psfig{figure=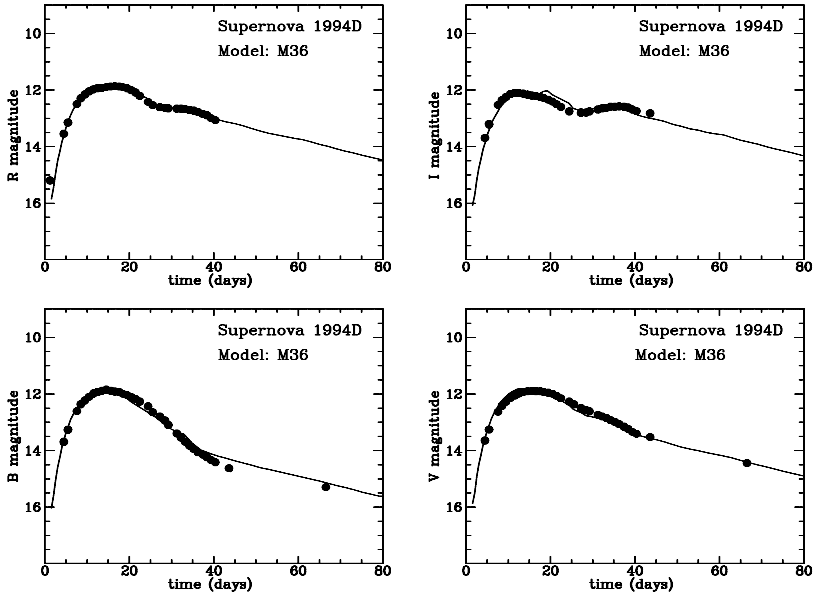,width=12.4cm,rwidth=7.9cm,angle=270}
\caption{Observed LCs of SN1994D in comparison to
the theoretical LCs of M36. \hfill }
\end{figure}
\begin{figure}
\psfig{figure=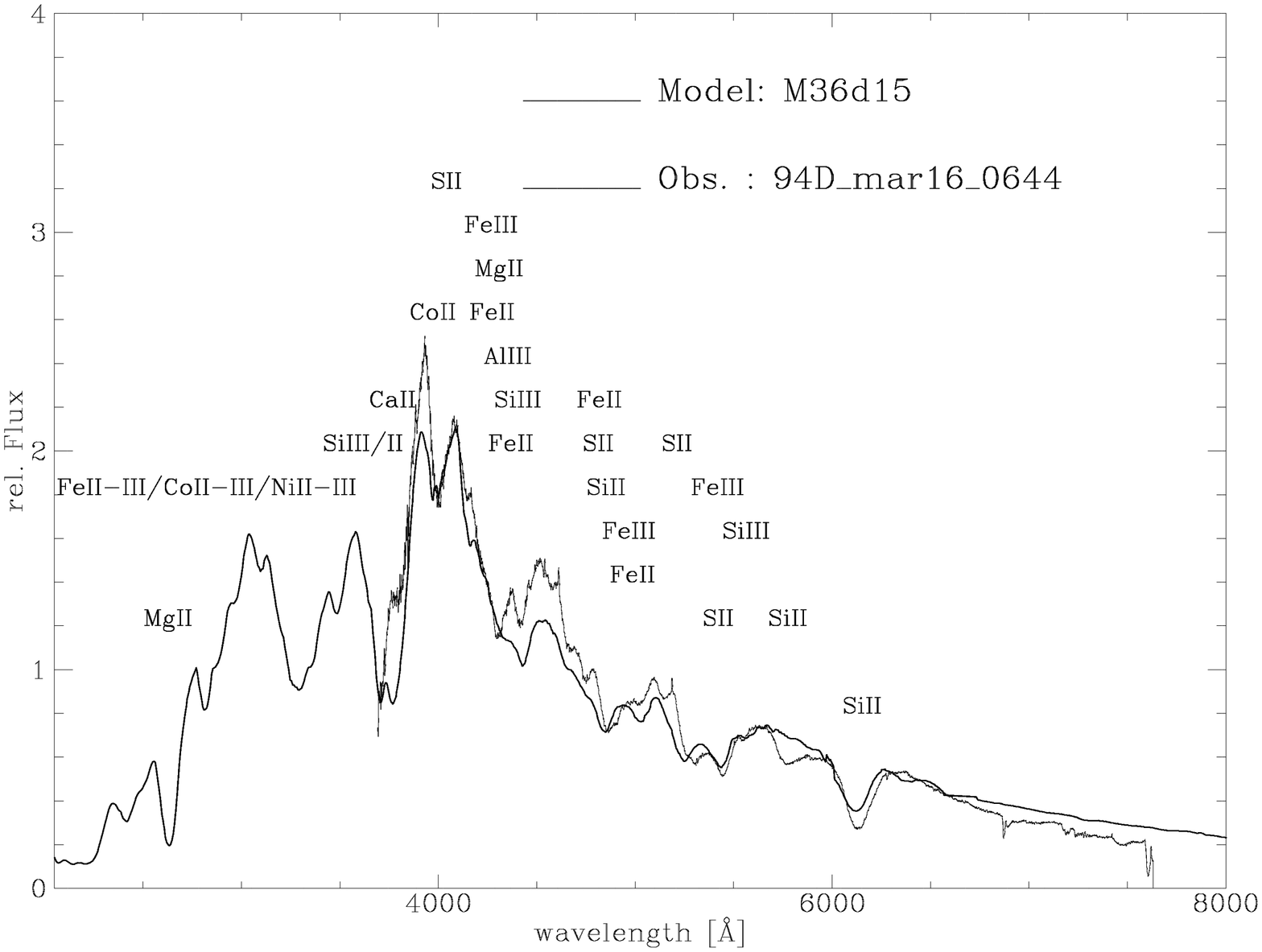,width=12.4cm,rwidth=6.5cm,angle=270}
\caption{Synthetic spectrum at day 15 for M36 compared to observations at Mar. 16th.}
\end{figure}

\noindent

 Although not presented in detail for all  models available (see Nomoto et al., 1984,
Khokhlov et al.  1993,  H\"oflich et al.  1994, H\"oflich, 1995, H\"oflich \& Khokhlov,
 1995),    we have compared  the observed optical  and infrared
 light curves of SN1994D for deflagration (W7, DF1),
 pulsating delayed detonation
 (PDD1a-c,3-9), delayed detonation (N21, N32, M35-39), helium detonations (HeD2-12),
 and envelope models (DET2env2-4) being 
 a crude representation of a merger scenario. 

 The pulsating delayed detonation models, helium detonation and envelope models
can be  ruled out because of the shape of their V and B light curves (Fig.1).
    W7 and N32 that  fit observations of several SNe~Ia                         
(M\"uller \& H\"oflich, 1994) provide better fits to SN1994D, but they fail to   
reproduce the IR light curves (H\"oflich 1995) and, less critical, the
post-maximum decline in V. One  may conclude that SN1994D is no
'standard' Type Ia supernova, but it is close to standard.
 We find best agreement between theory and observation for a the delayed detonation model
M36 with $\rho_{tr} = 2.4 ~10^{7} g/cm^3 $ and a deflagration
velocity of 0.03 times the sound velocity (Figs. 2 \& 3).
The initial central density of the WD
is $2.7~10^9 g/cm^{3}$, i.e. about 20 \% lower than in our delayed detonation models
previously considered. The lower density may be understood in terms of a higher
accretion rate  on the progenitor.   During the explosion, 
$0.6 M_{_\odot} $ of $^{56} Ni $ are produced.   
 Similar models with  $\rho_{tr} $
of $2.0 ~10^{7} $ and $3.0 ~10^{7} g/cm^{3}$ and  Ni productions of 0.51 and 0.67 
$M_\odot $, respectively, are not able to reproduce the observations.

From the vertical shift of the light curves, the time of explosion can be determined
to be between JD 244 9414.5 and 244 9415.5, meaning  SN1994D was discovered just about
3 to 4 days after the explosion!
 Our models are consistent with no interstellar reddening.
 Taking the uncertainties in our models  into account,
we determine the  distance of NGC 4526  to be   $  16 \pm 2  ~Mpc $.
The explosion took place between March 3rd and 4th, 1994.

 The NLTE-analysis of the spectra confirmed the results from the LC analysis and
give an insight to several new aspects.
Taking into account  that we do not allow for any artificial adjustment to provide
 better fits,  M36 reproduces the observed spectra reasonably well (Fig. 4),
including the evolution with time (H\"oflich, 1995). The agreement of the velocity shifts both
of the elements of partial burning such as Si, S, O and also the iron rich 
elements indicate  
that the chemical and density profiles of M36 seem to resemble SN1994D.  
 In the spectrum from March 11, the Doppler shift of the absorption minimum
of Si II indicates photospheric expansion velocities of $ \approx 15000 km/sec$ with
wings reaching out to more than 22000 km/sec.
 This strongly suggest the presence of a significant amount of Si at high velocities. 
Our  delayed detonation model M36
provide a Si mass fraction of 15 \% up to the outer layers ( $\approx 25000 km/sec$, 
 Fig. 1). For comparison, 
 the classical deflagration model W7 does not show Si at velocities $\leq 16400 km/sec$
because the smaller flame speed in the outer layers and, consequently, the larger
 pre-expansion of the WD envelope during the explosion. 
 In our previous analysis, we had to report the serious problem that the Ti features 
are too strong and, artificially, we had to reduce Ti in the outer layers.
 In  recent calculations for delayed detonation models (H\"oflich et al. 1995), we
find  that the Ti problem vanishes if we use a larger nuclear network and start with
a metallicity according to a population II star.

 Finally, we want to mention some of the limitations of our study.
   Despite similarities in both the spectra and light curves, the  photospheric 
expansion velocities inferred
from observed spectra show strong
individual variations (e.g. Branch 1987 and Barbon et al. 1990). Different models 
are needed 
to fit the light curves (M\"uller \&  H\"oflich  1994, H\"oflich et al.  1994). 
Therefore, the     results of our  study must not be generalized.
   Although we treat several elements in full NLTE,
detailed atomic models shall be implemented for  the iron group elements, namely Ti, Co, Ni.
 Our approach to the line  scattering by using an equivalent level approach for
the 'LTE-lines' seems
 to work,  but it must be regarded only as a first step towards a more consistent treatment. 
 Finally,  M36 produces only 0.003 $M_\odot $ of He in the region of 
$\alpha $ rich freeze-out whereas, a much larger amount is needed to produce a 
strong He feature at $1.\mu $. For a more detailed discussion of the latter problem,
 see Meikle  1995.
 
%
\parindent=0pt
\baselineskip=14pt
\section{References  }
 
\bibliographystyle{}
\begin{small}
 {Barbon R.,Bennetti S.,Cappalaro E.,Rosino L.,Turatto M.}  (1990),
 {\it A\&A}, {\bf 237}, {79}
\hfill
 
 {Branch  D.}  (1987), {\it ApJ}, {\bf 316}, {L81}
\hfill
 
{Canal R.}  ({1995}),{Les Houches Lectures},
eds. J. Audouze et al., Elsevier, in press
\hfill
 
 {H\"oflich  P. }  (1995), {\it ApJ }, {\bf 443}, 533
\hfill
 
 {H\"oflich  P., Khokhlov  A. }  (1995), {\it ApJ }, in press
\hfill
 
 {H\"oflich  P., Khokhlov  A., Wheeler  C. }  (1994), {\it ApJ }, 
{\bf 444}, 831
\hfill
 
 {H\"oflich  P., Khokhlov A., Nomoto  K., Thielemann F.K, Wheeler C.J.} (1995), this volume  
\hfill
 
 {Hoyle  P., Fowler }, ({1960}) {\it ApJ }, {\bf 132}, {565}
\hfill
 
 {Khokhlov  A., M\"uller  E., H\"oflich  P.} (1993), {\bf A\&A}, {\it
 270}, {23}
\hfill
 
 {Khokhlov  A.}, (1991), {\bf A\&A}, {\it 245}, {114}
\hfill
 
{Meikle P.} (1995), this volume
\hfill
 
 {M\"uller  E., H\"oflich  P.}, ({1994}), {\it A\&A}, {\bf 281}, {51}
\hfill
 
 { Nomoto  K., Yamaoka  H., Shigeyama  T., Iwamoto  K.}, ({1995}), 
{\sl in: Supernovae}, ed. R.A. McCray, {Cambridge University Press}, in press
\hfill
 
 {Nomoto K., Thielemann F.-K., Yokoi K.} ({1984}), {\it ApJ}, {\bf  286}, {644}
\hfill
 
 {Smith et al.}, {(1994)}, {\it ApJ}, {in preparation} {combined observations of CTIO and CfA}
\hfill
 
 {Treffers, Filippenko, Van Dyke} (1994), {IAU-circular} { 5946}
\hfill
 
{Wheeler  J. C., Harkness  R .P.}, ({1990}), {\it Rep. Prog. Phys.}, {\bf 53}{1467}
\hfill
 
{Woosley  S. E., Weaver T. A.}  ({1995}),{Les Houches Lectures},
eds. J. Audouze et al., Elsevier, in press
\hfill
\end{small}
 
 
\end{document}